\documentclass[journal]{IEEEtran}
\ifCLASSINFOpdf
\usepackage[pdftex]{graphicx}
\else
\fi

\begin{document}
\title{The Atacama Large Millimeter/submillimeter Array}

\author{Alwyn~Wootten and A.~Richard~Thompson, Life Fellow IEEE}% <-this % stops a space
\thanks{A. Wootten and A. R. Thompson are with the National Radio Astronomy Observatory, 520 Edgemont Road, Charlottesville, Virginia 22901  USA e-mail: (awootten@nrao.edu, athompson@nrao.edu).}% <-this % stops a space
\thanks{Manuscript last draft March 26, 2009.}


\markboth{Proceedings of the IEEE}{}%
%{Shell \MakeLowercase{\textit{et al.}}: Bare Demo of IEEEtran.cls for Journals}



\maketitle

\begin{abstract}
%\boldmath
The Atacama Large Millimeter/submillimeter Array (ALMA) is an international radio telescope under construction in the Atacama Desert of northern Chile.  ALMA is situated on a dry site at 5000 m elevation, allowing excellent atmospheric transmission over the instrument wavelength range of 0.3 to 10 mm.  ALMA will consist of two arrays of high-precision antennas.  One, of up to 64 12-m diameter antennas, is reconfigurable in multiple patterns ranging in size from 150 meters up to $\sim$15 km.  A second array is comprised of a set of four 12-m and twelve 7-m antennas operating in one of two closely packed configurations $\sim$50 m in diameter.  The instrument will provide both interferometric and total-power astronomical information on atomic, molecular and ionized gas and dust in the solar system, our Galaxy, and the nearby to high-redshift universe.  In this paper we outline the scientific drivers, technical challenges and planned progress of ALMA.  
\end{abstract}

% \fi
%\IEEEpeerreviewmaketitle
\section{Introduction}
In the total electromagnetic spectrum of the Universe,
there are three major peaks.  One, the biggest, is the peak from the 3
K blackbody radiation relic of the Big Bang.  That peak
occurs in the millimeter wavelength range of the spectrum, as expected for any
black body radiating at such a low temperature.  The third strongest peak occurs
near one 1 micron (1 $\mu$m) wavelength: this contains the accumulated light from all of the stars and
planets in the Universe.   The second strongest occurs at about 1.5 THz or 200 microns wavelength.  Light near this wavelength cannot penetrate the atmosphere, as it is absorbed
by water and other molecules in the atmosphere: this peak was identified
only recently through satellite observations.  This  spectral feature represents all of the cool ($\sim$200 K)  objects in the Universe, that is, clouds of dust and gas as well as radiation from warmer sources that is absorbed and reradiated.  Alas, with a satellite
one is limited as to the size of telescope one can observe with and hence the resolution obtained.
Current spacecraft apertures are far too small to give good
images of what produces this second peak.  ALMA\footnote{The Atacama Large Millimeter/submillimeter Array (ALMA), an international astronomy facility, is a partnership of Europe, Japan and North America in cooperation with the Republic of Chile. ALMA is funded in Europe by the European Organization for Astronomical Research in the Southern Hemisphere, in Japan by the National Institutes of Natural Sciences (NINS) in cooperation with the Academia Sinica in Taiwan and in North America by the U.S. National Science Foundation (NSF) in cooperation with the National Research Council of Canada (NRC) and the National Science Council of Taiwan (NSC). ALMA construction and operations are led on behalf of Europe by ESO, on behalf of Japan by the National Astronomical Observatory of Japan (NAOJ) and on behalf of North America by the National Radio Astronomy Observatory (NRAO), which is managed by Associated Universities, Inc. (AUI).}, with excellent sensitivity and resolution at a high dry location will allow sensitive imaging in the range 31-950 GHz (wavelength range of approximately 1 cm to 300 $\mu$m).  Thus ALMA will observe within the wavelength regimes of the strongest two radiation peaks, to the extent that Earth's atmosphere allows.

ALMA consists of two parts.  There is an array of 12 m diameter antennas, the scientific requirement for 64 of which will ensure full realization of  the scientific goals set forth in the Bilateral Agreement;  contracts in place will provide at least 50 antennas.  For this array baselines extend from 15 m to $\sim$15 km.  We refer to this antenna complement as the ``12-m array''.  There is also the Alma Compact Array (ACA) which consists of four 12 m antennas plus twelve 7 m antennas \cite{Iguchi2009}.   The smaller diameter of the 7 m antennas allows a minimum antenna spacing of 8.75 m.  A summary of ALMA specifications can be found in Table I.  

\begin{table}[ht]	%\label{spectable}
\caption{Summary of ALMA Specifications}
\begin{center}
\scriptsize
\begin{tabular}{lrr}
\noalign{\medskip}
%\tableline
\noalign{\smallskip}
Parameter & 12m Spec & 7m Spec \\
\hline
\noalign{\smallskip}
Number of Antennas & up to 68 & 12 \\
Antenna Diameter & 12 m  & 7 m\\
Antenna primary focal ratio (f/D)$^a$ & 0.4 & 0.37\\
Geometrical Blockage & $<$3\% & $<$5\% \\
Antenna Surface Precision & $<$ 25 $\mu$m rms & $<$ 20 $\mu$m rms \\
Antenna Pointing Accuracy & $<$ 0."6 rms & $<$ 0."6 rms \\
Total Collecting Area & 6600-7700 m$^2$ & 462 m$^2$ \\
Antenna primary beam &  17" x $\lambda ^b$ (mm) & 30" x $\lambda$ (mm) \\
Max (finest) Angular Resolution & 0.015" x $\lambda$ (mm) & 5" x $\lambda$ (mm) \\
Configuration Extent & 150 m to 14 km &  41 m \\
Correlator Bandwidth & 16 GHz per baseline & same\\
Spectral Channels & 4096 per IF  & same \\
Number of 2 GHz-wide IFs & 8 & same\\
%\tableline
%\tableline
\end{tabular}
\end{center}
$^a$f indicates focal length, D indicates primary diameter.
$^b\lambda$ indicates wavelength.
\end{table}

ALMA has three primary science goals, defined in the Bilateral Agreement by which the observatory was founded (for a history of ALMA, see \cite{vandenBout2005}; for a compendium of science, see \cite{Wootten2001} and \cite{ALMA2008}).  The first of these goals is to detect emission from the CO molecule or C$^+$ ion towards a galaxy of Milky Way luminosity at a redshift of 3 (see discussion in \cite{deBreuck2005}) in less than 24 hours integration.  Although CO and C$^+$ lines have been detected in more distant galaxies, those galaxies emit a much higher luminosity than the Milky Way.  The sensitivity needed for this measurement sets the collecting area of ALMA, since sensitivity cannot be increased by enlarging the bandwidth for spectral line detections.   It also drives the receiver specifications, discussed below.  The current construction scope of the project will provide at least 6600 m$^2$ of collecting area.  A second top-level science goal (see discussion in \cite{Richer2005}) requires ``The ability to image the gas kinematics in protostars and protoplanetary disks around young Sun-like stars at a distance of 150 parsecs (pc), enabling one to study their physical, chemical and magnetic field structures and to detect the gaps created by planets undergoing formation in the disks''.  This requirement drives angular resolution, since a disc of radius equal to Jupiter's orbit, for example, subtends only 0.03" at 150 pc.  To meet this requirement, ALMA's resolution must be finer than this.  To measure the chemical properties of the disk, one needs a wide bandwidth, capable of simultaneously measuring emission lines from many different atomic or molecular species.  To measure magnetic fields, the system must be able to measure the resulting polarized components of the incoming waves.  The third high-level goal is to provide excellent imaging, which requires an optimized distribution of the antennas and correction for imaging errors, the most severe of which are induced by turbulence in the Earth's atmosphere.  Other science goals drive other aspects of ALMA instrumentation, and together the goals have led to the scientific and technical requirements of the instrument under construction today.

\section{The Chajnantor Site and the Array Configuration}

In May 1998 NRAO recommended construction of ALMA's conceptual predecessor, the Millimeter Array, on a site in Region II of northern Chile which lies on a plain at the foot of three volcanic peaks, Cerro Toco,
%Question Mark.  Also put ref to memo on the joint dist of phase and opacity.  Memo 521.
Cerro Chajnantor and Cerro Chascon.  This area had also been investigated by European and Japanese colleagues.  The site 
(longitude 67$^o$ 45' W, latitude 23$^o$ 01' S) lies just north of the Tropic of
Capricorn,  about 50 km east of the historic village of San Pedro de Atacama, 130 km southeast of the mining
town of Calama, and about 275 km ENE of the coastal port of Antofagasta. 
It lies close to the border with Argentina and Bolivia beside the Paso
de Jama road into Argentina.  The mean elevation is about 5000 m (16400 ft).
Testing of the Chajnantor site began
in April 1995 and continued until 2005, a joint effort of NRAO and
the European Southern Observatory (ESO).  The Nobeyama Radio Observatory
(NRO) operated a similar testing facility nearby at Pampa la Bola.  

\begin{figure}[!h]
\centering
\includegraphics[width=3.5in]{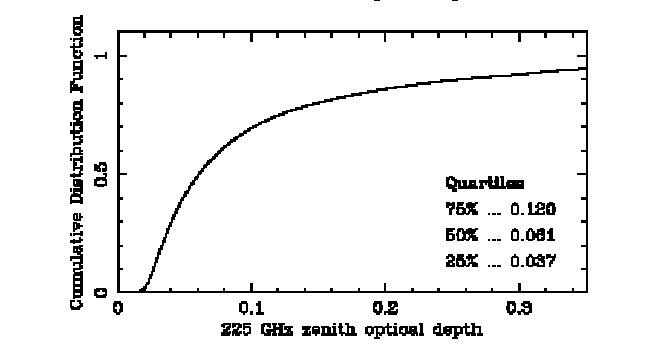}
\caption{Cumulative distribution of 225 GHz optical depth at Chajnantor measured over the period 1995-2004. }
\label{fig_sim}
\end{figure}

\begin{figure}[!h]
\centering
\includegraphics[width=3.5in]{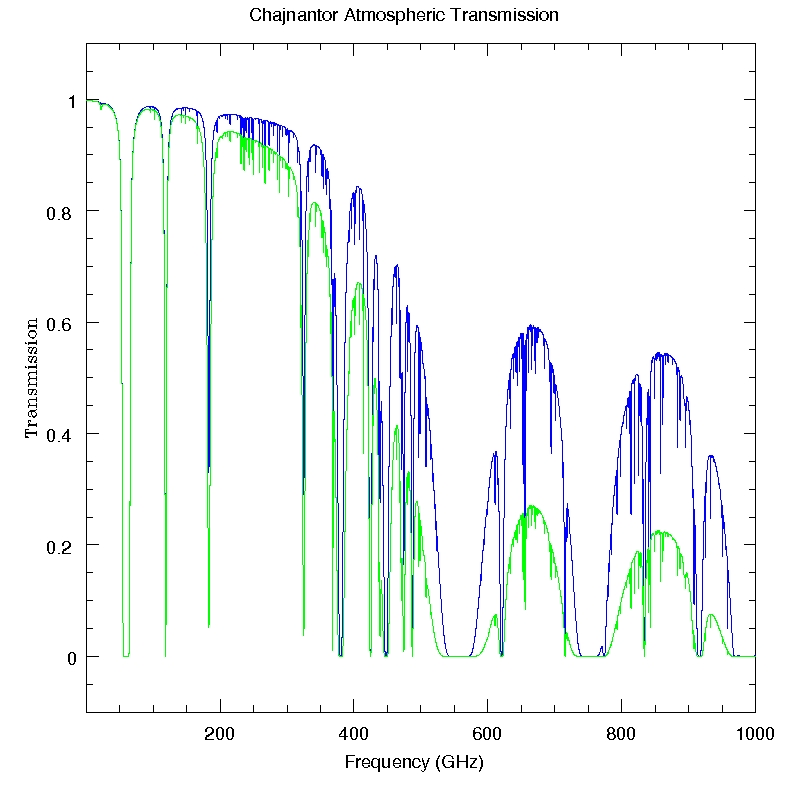}
\caption{Zenith transmission of the atmosphere at the Chajnantor plain for the median atmospheric content of precipitable water vapor (green; 1.26 mm) and for the best octile  (blue; 0.47 mm), from model predictions.}
\label{fig_xm}
\end{figure}

Some salient characteristics from the testing carried out at the site include: the median annual temperature is -2.5$^o$C with annual 50th percentile winds of 10.4 m/s.  The average barometric pressure for this elevation is only 55 percent
of the value at sea level.  Humidity averages 39\% and ultraviolet radiation is about 170\% of that at sea level.  Transparency at 225 GHz has been monitored for several years \cite{Radford2002}; the 50th percentile zenith optical depth at this frequency is 0.061, which using atmospheric modeling,  corresponds to a column of precipitable water of a little more than
1 mm.  The cumulative distribution of opacity measured at 225 GHz over that period is shown in Fig. \ref{fig_sim}.   The joint distribution of opacity and phase is discussed in \cite{Daddario2005}.  With such a low water column, observations are possible at the atmospheric windows covered by the two highest ALMA Bands (602-950 GHz) for
roughly half of the time (see Fig. \ref{fig_xm} and see Table II).  Even the super THz windows at 1.035 and 1.3 to 1.5 THz have been measured from Chajnantor \cite{Peterson2003} showing transmission of up to 30\%.  In fact, the APEX telescope \cite{Gusten2006}, a modified version of the Vertex design for the ALMA prototype, has operated on the site since 2005 and has observed in the 1.5 THz window \cite{Wiedner2007}.

\begin{figure*}[!t]
\centering
\includegraphics[width=7.0in]{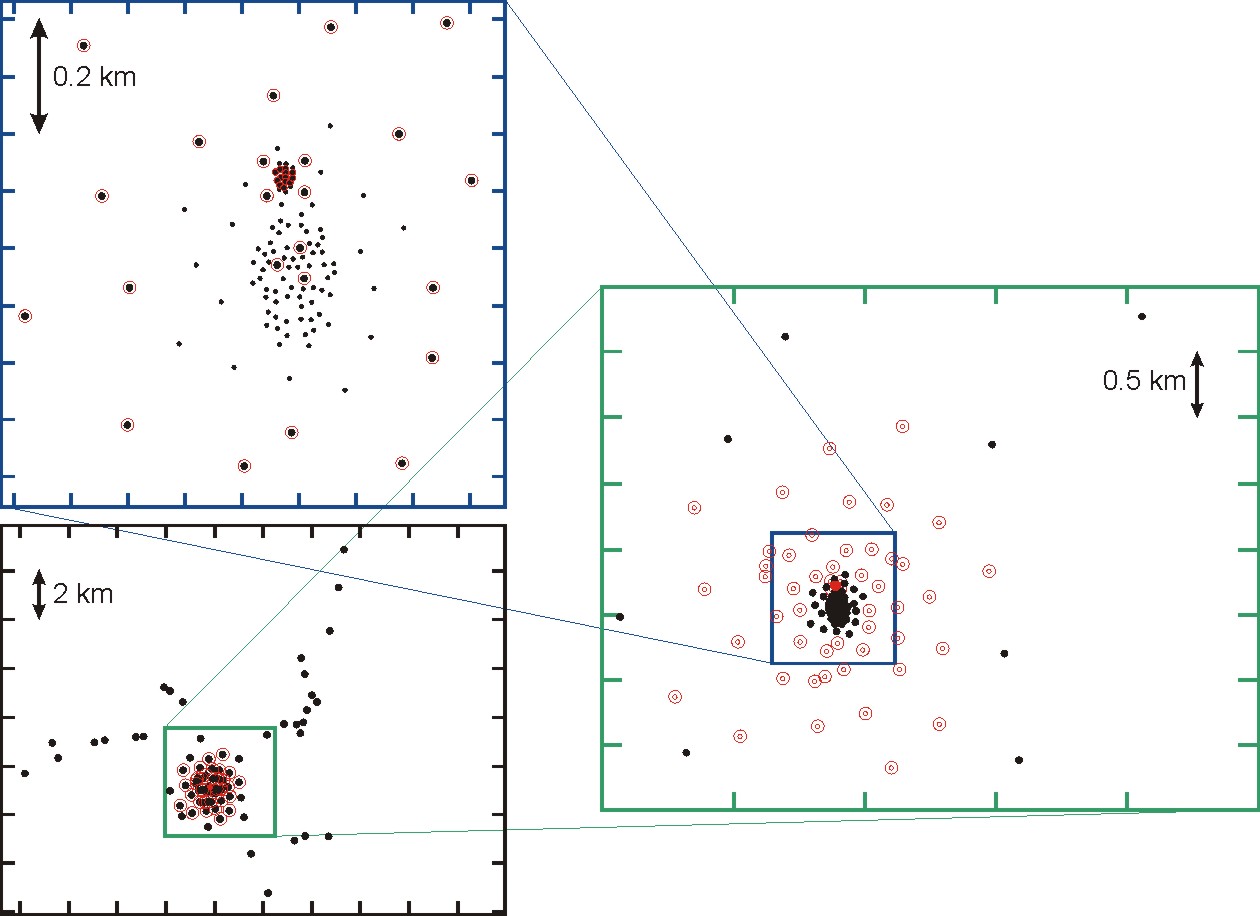}
\caption{Antenna foundation locations are shown on three different scales, from lower left on the largest scale, to right on an intermediate scale and on the upper panel in a closeup of the central compact configuration and the ACA.  Antenna foundations in red highlight those occupied by an antenna in configuration 18, which is appropriate for a circular beam on a source transiting at high elevation angles.  Note that three antennas remain on short baselines, to provide short spacings and spatial dynamic range to images.  In the lower left panel the locations of the outer antennas are determined by several factors in addition to optimization of imaging quality: the ground contours, minimization of road/power/fiber costs and avoidance of wildlife habitat.  ACA antennas (small dense group clustered above the center of the upper figure) do not normally move upon reconfiguration but may be reconfigured to provide nearly circular beams at extreme declinations.  }
\label{fig_config}
\end{figure*}

 As the array is reconfigurable, each of the antennas must be transportable on an occasional basis by a transporter.  Two of these massive vehicles have been designed and built by Scheuerle Fahrzeugfabrik GMBH and currently operate on the ALMA site (Fig. \ref{fig_transporter}).  Antennas are transported between the Operations Support Facility (OSF) at 2900 m, where they are assembled and maintained, and the Array Operations Site (AOS) at 5000 m where the array is located.  During normal operations observations will be controlled from the OSF.  At the AOS there is a set of the nearly 200 foundations (Fig. \ref{fig_config}) on which the antennas can be mounted, to and from which signals travel by optical fiber between the antennas and the AOS Technical Building.  The transporters move antennas between the foundations according to the observing demand.  Moving through a complete cycle of configurations, from the most compact to the most extended can take several months.  The maximum baselines available are on the order of 15 km.  Preliminary plans are to move antennas on a few-per-day basis along a self-similar configuration of roughly spiral geometry out to the largest configuration, where topography heavily constrains the design.  The inner 4 km foundation positions are based on a single-arm, tightly-wound spiral, with a ratio in radial distance between adjacent antenna stations of 1.041 and change in azimuth angle of 138.97 degrees.  The plan involves approximately 28 steps as the configuration is expanded from the most compact to the most extended form.  For each each expanding step the beam size of the configuration decreases by 17\%.  A set of alternate configurations provides beams optimized for high elevation or for extreme north or south declination targets.  This will provide images with a range of detail; any given configuration will provide excellent imaging.  A configuration example is shown in Fig.  \ref{fig_config}.  The ACA antennas occupy a special set of closely spaced foundations which allow for limited repositioning to improve the resolution for extreme north or south declinations.

%\begin{figure}[!h]
%\centering
%\includegraphics[width=3.0in]{/Users/awootten/Desktop/ALMA/almasci/ALMAPapers/IEEESpecialIssue/IEEEtran/ARThompson/Config18uv.pdf}
%\caption{Outward configuration number 18.  Blue dots show transit snapshot {\it u,v} coverage at zenith
%angle of 23$^\circ$ (corresponding to declination 0 or -46$^\circ$). Red dots show antenna
%positions.}
%\label{fig_config}
%\end{figure}

\section{The Antennas}

\begin{figure}[!b]
\centering
\includegraphics[width=3.0in]{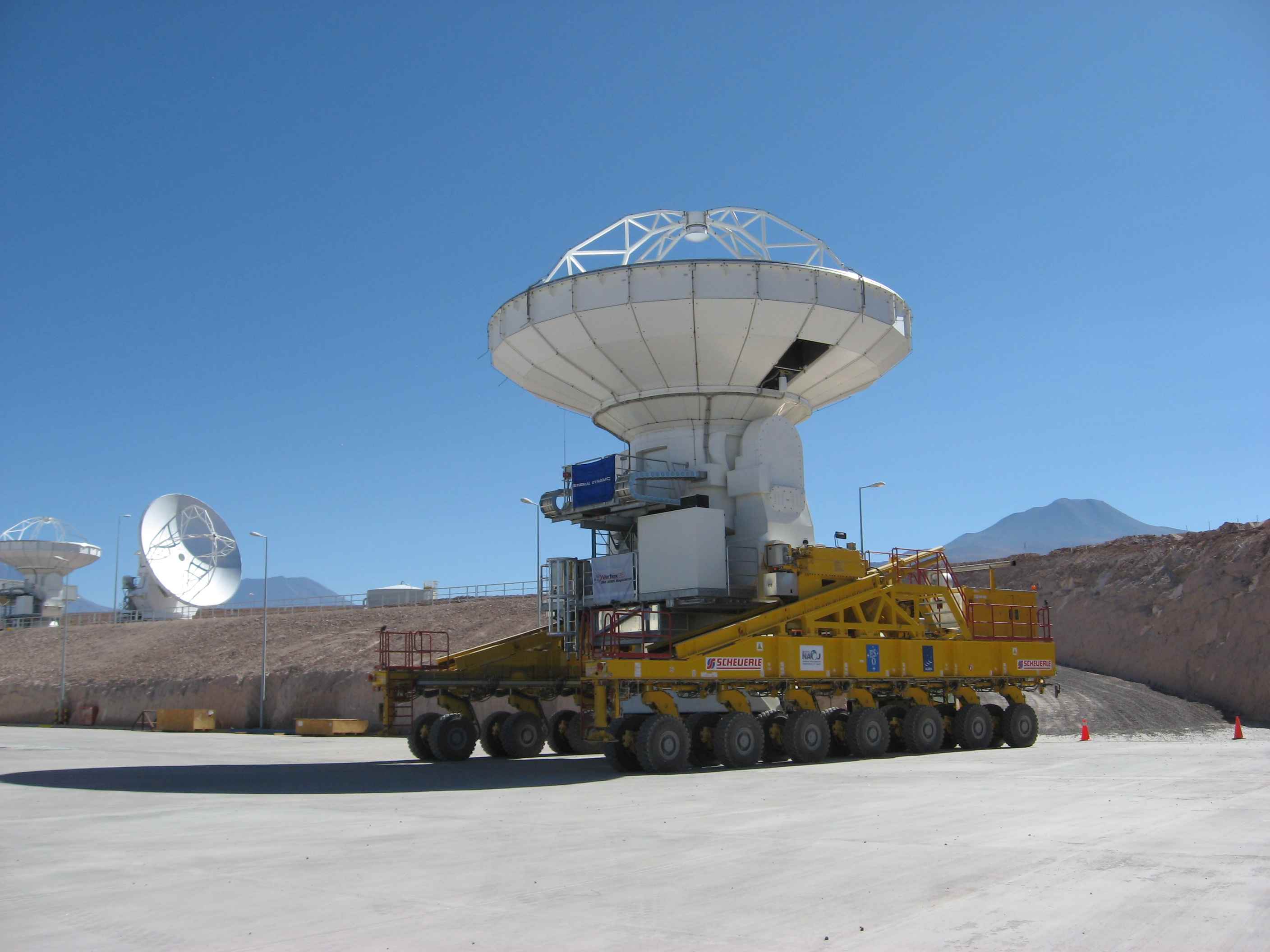}
\caption{The first antenna accepted by ALMA from Vertex is moved to a pad near the Operations Support Facility Technical Building on one of the transporters.  In the background are the first antenna accepted by ALMA, from Melco, undergoing holography tests on the right, and Vertex No 4 on the left.}
\label{fig_transporter}
\end{figure}

For both the 12-m array and the ACA, all of the antennas have alt-azimuth mounts and a quadripod mounting for the subreflector. The ACA is intended for studies of more extended objects, and the four 12 m antennas will be equipped for total-power measurements (particularly continuum), that is, single-antenna measurements which provide the zero-baseline  values of the fringe visibility.  The subreflectors of these four antennas can be moved so that the beam jumps rapidly between two positions a few beamwidths apart on the sky, to measure the signal power on-source and off-source.  The ACA can be used in combination with the 12-m array or as an independent instrument.  

\begin{figure}[!t]
\centering
\includegraphics[width=3.0in]{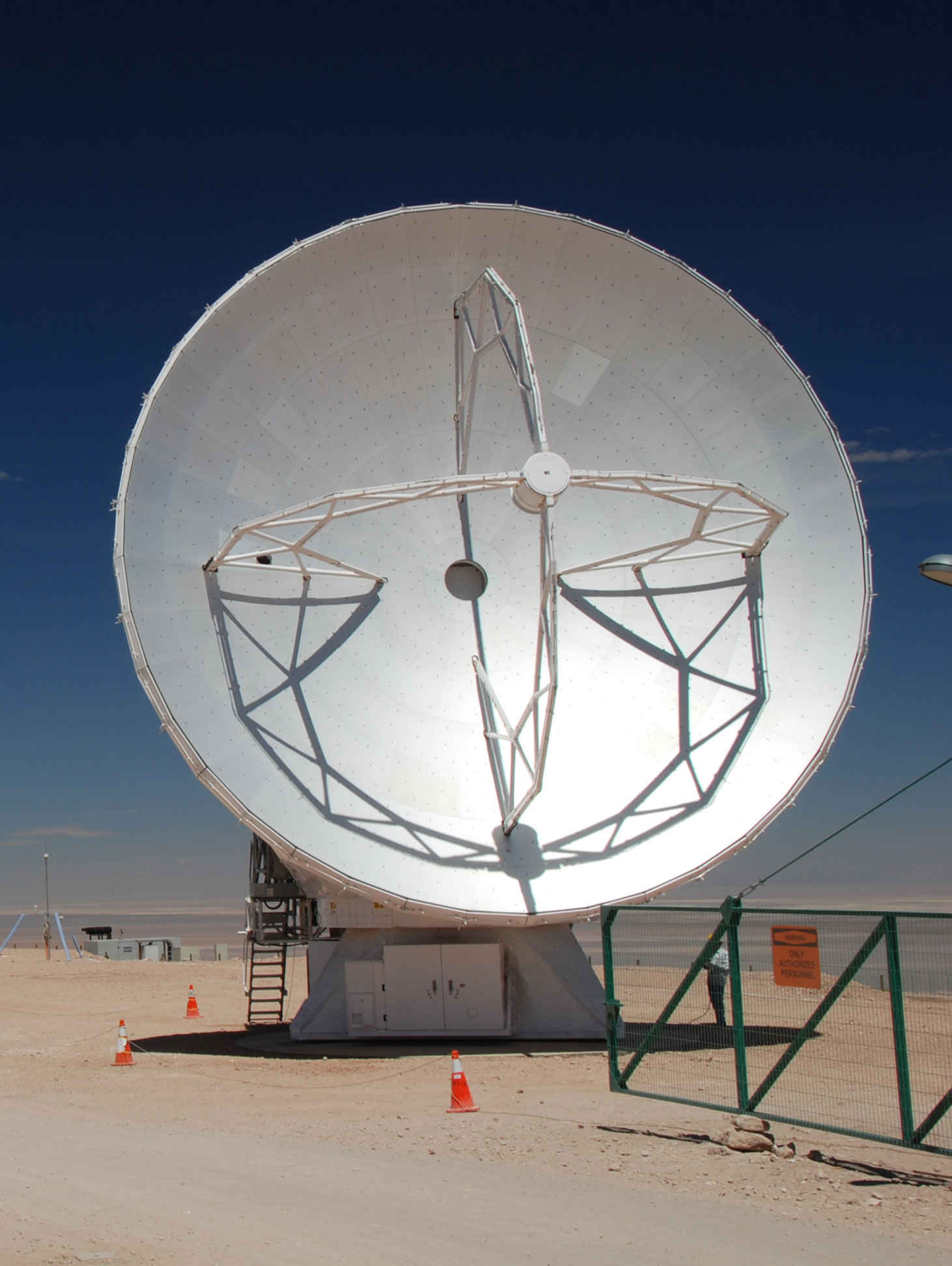}
\caption{The first 12 m antenna accepted by ALMA from  Melco. }
\label{fig_Melco}
\end{figure}

Each of the partners, NRAO, ESO, and NAOJ, contracted for construction of a prototype antenna that would meet specifications derived from the scientific goals of ALMA.  These prototype antennas were
tested at the NRAO's Very Large Array (VLA) site near Socorro, N.M \cite{Mangum2006}\cite{Snel2007}\cite{Baars2007}\cite{Ukita2004}.   Subsequent to the prototype testing phase, antenna production contracts were written in 2005 with Vertex, AEM and MELCO.  Vertex (Fig. \ref{fig_transporter}) is part of the 
%General Dynamics SATCOM Technologies division of the 
General Dynamics Corporation of the USA, AEM is a European consortium consisting of Thales-Alenia Space, European Industrial Engineering and MT Mechatronics and MELCO is part of the Mitsubishi Electric Corporation of Japan (Fig. \ref{fig_Melco}).  The designs make use of advanced techniques to achieve ALMA's demanding requirements. 
The ALMA prototype and production antennas make extensive use of carbon fiber reinforced plastic (CFRP) technology, in order for the antenna to maintain a stable parabolic shape in the harsh thermal (temperatures from -20$^\circ$ to +20$^\circ$C) and wind  environment characteristic of the ALMA site.  The antenna surface accuracy must be better than 25 microns to enable efficient observations at the very highest frequency.  The antennas will also be used to observe the Sun, and the panel surfaces have been specially treated to diffuse infra-red and shorter wavelengths.  Much of the time, ALMA will image fields larger than the primary beam, which means that multiple pointings of the antenna will be combined to produce a single image.  This requires that the antennas maintain an offset pointing accuracy better than 0.6" , despite breezy conditions (6 m/s (day); 9 m/s (night)).  The Chajnantor site affords no vegetative cover of consequence, so windblown grit and dust will occur and must not degrade the performance of the antenna.   
%The ultraviolet radiation at this altitude is 170\% of that at sea level.  
All these factors provide a challenge to modern antenna design, which testing of the prototype antennas has shown to have been met.

The antennas use Cassegrain optics.  Since the feeds are slightly ($<$0.5 m) off-axis in the ALMA design (see Section IV), the 0.75 m subreflector is tilted to focus the signal onto the one currently in use.  A 60 mm diameter cone is machined in the center of the subreflector in order to obtain good suppression of the power reflected from the central area \cite{Hills2005}.  A similar cone has also been used on the IRAM 30-m millimeter radio telescope \cite{Baars1994}.  Reflection back into the focal plane due to diffraction from the outer edge of the subreflector is suppressed by using  a rounded-off edge.  Analysis of effects due to blocking and scattering can be found in \cite{Morris1978}.

\section{The ALMA Receiving System}
\begin{figure}[!t]
\centering
\includegraphics[width=3.0in]{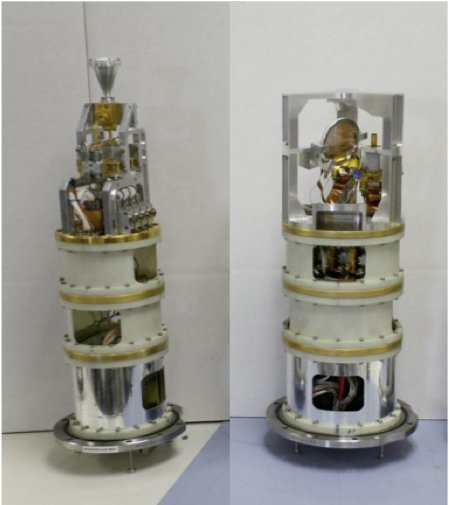}
\caption{Two of the cartridges for the ALMA Front End dewar.  Left:  Band 3 (84-116 GHz) cartridge.  Right:  Band 7 (275-373 GHz) cartridge.}
\label{fig_cartridge}
\end{figure}

%\end{center}
Each of the 12 m ALMA antennas will be equipped with a receiving system to cover the full range 31-950 GHz, in ten bands as listed in Table II.
%\ref{receivers}.
Each receiver is packaged in the form of a cylindrical cartridge (see e.g. \cite{Ediss2004}\cite{Satou2008}) with the feed horn at the top (Fig. \ref{fig_cartridge}).  The receiver cartridges are mounted within a circular Dewar of radius approximately 1 m, the top surface of which is just below the vertex of the antenna.  For each feed horn there is a circular microwave window in the top of the Dewar, to admit the radiation from the antenna subreflector.  Within each receiver cartridge the output from the  feed is separated into two linearly polarized components by an orthomode transducer, or in the case of the four highest bands by a quasi-optical grid system.  The Dewar allows parts of the receiving systems to be held at temperatures of $\sim$100, 15, or 4 K, as required for different components.

\begin{table}[h]	\label{receivers}
\caption{Summary of ALMA Receivers}
\begin{center}
\scriptsize
\begin{tabular}{lccc}
\noalign{\medskip}
%\tableline
\noalign{\smallskip}
Band & Frequency & T$_{SSB}^a$ &  Configuration\\
 & (GHz) & (K) &  \\
\hline
\noalign{\smallskip}
1 & 31 - 45 & 17  & HEMT\\
2 & 67 - 90 & 30 & HEMT\\
3 & 84 - 116 & 41  & 2SB\\
4 & 125 - 163 & 51  & 2SB\\
5 & 163 - 211 & 65    & 2SB\\
6 & 211 - 275 & 83 &  2SB\\
7 & 275 - 373 & 147 &  2SB\\
8 & 385 - 500 & 196 &    2SB\\
9 & 602 - 720 &175$^b$ &   DSB\\
10 & 787 - 950 & 230$^b$ &  DSB\\
%\tableline
%\tableline
\end{tabular}
\end{center}
$^a$Requirement for 80\% of the radio frequency band.
$^b$DSB receiver noise temperature is given for Bands 9 \& 10.
\end{table}

For each frequency band the receiver is designed to provide the lowest possible noise temperature.  For the two lowest frequency bands the input stages are high electron mobility transistors (HEMT) amplifiers
cooled to 15 K.  For Bands 3 and higher, SIS (superconducter-insulator-superconductor) tunnel-junction mixers, requiring a physical temperature of 4 K, are the optimum choice.  However, with mixers as the first stage, there is a response to both the upper and lower (or signal and image) sidebands, and removal of one of these by filtering at the input is not possible without introducing some loss.  The best solution is the use of sideband-separating mixers, in which the input signal is divided into two components with $90^\circ$ relative phase and applied to two different mixer elements.  The IF outputs of these mixers are amplified and then combined in a quadrature hybrid.  The signals from the two sidebands appear at different output ports of the hybrid.  Systems of this type are referred to as dual sideband (2SB) and are used for bands 3-8. To minimize losses, compact designs have been developed in which the RF hybrid and a local oscillator (LO) power divider are in the form of waveguide circuits in a single split metal block, within which all the elements of the 2SB mixer are mounted, see, e.g., \cite{Kerr2004}, \cite{Chin2005} or \cite{Claude2008}.  As the wavelength becomes shorter, it is increasingly difficult to produce such compact designs and above 600 GHz, for the top two bands, double sideband (DSB) mixers without sideband separation are used, see, e.g., \cite{Shan2007}.  In the 2SB receivers the unwanted sideband rejection is in some cases no more than $\sim$10 dB.  This is sufficient to make use of the greater sensitivity of 2SB systems relative to DSB.  To further reduce spectral features from the unwanted sideband, additional suppression is available using 90$^\circ$ phase switching (see section VII), or such features can be identified by introducing a frequency offset in the first LO.  No mechanical adjustments are required in the operation of the receiving system.  In each receiver unit there are two sets of amplifiers or mixers to accommodate the two polarized components of each signal.

\begin{figure}[!b]
\centering
\includegraphics[width=3.0in]{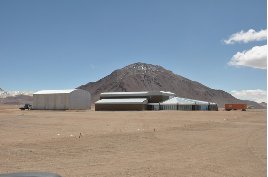}
\caption{The Array Operations Site Technical Building houses the correlator and the Local Oscillator System.  At 5000 m elevation, it remains unstaffed most of the time.}
\label{fig_AOS}
\end{figure}

The intermediate frequency (IF) stages that follow the input stages described above provide a total bandwidth of 16 GHz.  The IF amplifiers incorporate HEMTs at 15 K temperature.  For bands 3-5 and 7-8, for which there are separate outputs for each sideband, the 16 GHz bandwidth provides 4 GHz per sideband and per polarization.  For Band 6 (1.3mm) a full 8 GHz of bandwidth per polarization in a particular sideband is processed.   The IF amplifiers have responses in the range 4-12 GHz, and their outputs are divided into 2 GHz-wide bands for digitization.  Three-bit quantization is used with 4 gigasamples/s rate\cite{Recoquillon2005}.  This results in a total bit rate of 96 Gb/s for each antenna.
These digitized data are transmitted by single-mode optical fiber from the antenna to the AOS Site Technical Building (Fig. \ref{fig_AOS}) where the correlator is located.

\section{Digital Back-End and Correlator}

Before cross correlation of the signals is performed, compensation must be made for the varying time delays resulting from the geometrical paths from the incoming wavefront to the individual antennas, as well as for differences in the transmission paths between the antennas and the correlator location.  The locations at which the antennas are mounted are precisely known from observation of radio sources with very accurately known positions.  Thus the geometrical delays for an incoming wavefront from the center of the field of study are accurately calculable.  The relative timing of the signals from the different antennas is coarsely adjusted in increments equal to the sampling interval by the use of Random Access Memory (RAM) delay lines.  Fine adjustment to $\sim 1/16$ of the sampling interval is effected by inserting variable time offsets in the digital sampler clocks \cite{Thompson2006}.  

The ALMA correlator \cite{Escoffier2007} for the 12-m array is designed to process all cross products for up to 64 antennas, with total bandwidth 16 GHz per antenna.  The correlator is a development of the XF type, i.e.\ cross multiplication with time lags followed by Fourier transformation to provide frequency spectra.  In the ALMA correlator, these processes are preceeded by digital filtering which divides each 2 GHz-wide band into 32 sub-bands of width 62.5 MHz.  As a result, the spectral resolution is a factor of 32 times greater than that provided by the range of time lags in the cross correlation.  A further factor of two in the resolution is available by digital filtering to 32 bands of width 31.25 MHz, but with a reduction of a factor of 2 in the total bandwidth covered.  The center frequency of each filter is independently tunable, thus providing high flexibility to the correlator system.  The initial filtering into sub-bands can be bypassed, in which case the correlator behaves as a pure XF system with faster dump rates but with much less spectral resolution.  At the filter outputs the signals are requantized to 2 bit (four level) or 4 bit (16 level) for cross-correlation.  The multiply-and-add operations that follow are performed at a clock rate of 125 MHz (i.e. the 4 GHz sample rate divided by 32), and with a full complement of 2016 baselines in use the number of such operations per second is $1.7\times 10^{17}$.  Cross correlations between orthogonal as well as parallel planes of polarization of the received signals are performed, so that images of any polarized component of the brightness distribution can be made.  Cross correlation data go to a long term accumulator and are integrated for integer multiples of 16 ms.  Autocorrelation (total power) data are integrated for multiples of 1 ms.  The ALMA correlator is constructed in the form of four ``quadrants'' each of which processes the full 2016 baselines with both polarizations for 4 GHz of bandwidth.  Each quadrant occupies eight equipment racks.  The total estimated power dissipation of the full correlator is around 170 kW.  It will be possible for instances in which maximum sensitivity is needed  to process signals from ACA antennas in the 12-m array correlator.  For example this benefits bandpass calibration for the 7 m antennas as it permits them to be correlated with the ensemble of 12 m antennas rather than with those in the ACA alone.

The ACA antennas can be used with the 12 m antennas with the correlator described above, or more often the ACA can operate using its own ACA correlator \cite{Iguchi2009}.  The ACA correlator \cite{Okumura2002} adopts an FX method in which the hardware calculates the fast Fourier Transform of data from each antenna, and then performs multiplications in the frequency domain.  The input FFT has an effective length of 524,288, producing a spectral bandwidth of 3.8 kHz across the 2 GHz IF channel.  These data are processed so that the ACA correlator modes match those of the 12 m array correlator.  For the ACA, fringe rotation is performed in the F-part of the ACA correlator except when the ACA is operating with the 12 m correlator, in which case fringe rotation is performed at the antennas.   

\section{Correction for the Atmosphere}
At the very short radio wavelengths at which ALMA operates, it is essential to make corrections for the effect of phase errors induced by variations in the atmospheric path delay.  These can result from water vapor and from variations in the density of the dry component of the atmosphere.  Variations in the water vapor are monitored by measurement of the intensity of the radiation from the atmosphere in the 183.31 GHz water line \cite{Hills2001}.  Each 12 m antenna is equipped with an additional receiver and feed for this measurement.  The width of the beam at 183 GHz is $\sim$33".  The receiver uses an uncooled Schottky-diode mixer as a double-sideband input stage, and the input is rapidly switched to measure the difference in noise temperature between the sky and a cold load\footnote{At the time of writing the testing phase of the atmospheric correction system is in progress, and the choices of techniques and parameters are not yet final.}.  The receiver passband covers approximately 175-191 GHz and contains four filters that provide measures of the peak value and width of the line \cite{Hills2007}.  These data are combined with other weather information to calibrate an atmospheric model that provides corrections to the observed visibility phases about once per second.  The water line data are insensitive to the dry component of the atmosphere, which can be monitored by moving to a nearby calibration source at intervals of about five minutes.  This provides a calibration of the instrumental and dry atmospheric phase components.  

\section{Local Oscillator (LO) System}

The major technical problem of the LO system is maintaining sufficient phase stability in the LOs used in the first frequency conversion in the receiver front ends.  The main LO system is located in the AOS Technical Building, and the required first LO frequency, or a subharmonic, is distributed to each antenna by optical fiber \cite{Shillue2004}\cite{Cliche2004}.  This outgoing frequency is in the range 27-142 GHz and a final frequency multiplication by a factor of 3, 5, or 9 is required at the antenna for the higher frequency bands.  The outgoing frequency is transmitted as the difference between two laser frequencies, a master laser and a slave laser which is offset from the master by the frequency required at the antenna\footnote{A possible alternative \cite{Kiuchi2007} to producing the two frequencies in different lasers is by modulation of a single laser using a Mach-Zender system which produces sidebands at (laser frequency)$\pm$(modulation frequency) while suppressing the laser frequency.}.  The offset frequency is phase locked to the system master oscillator to maintain the phase of the laser difference frequency at the antenna.  There, it is used to phase-lock a signal generated by a YIG-tuned oscillator and multiplier chain, which drives the final frequency multiplication for the higher bands performed by a cold ($\sim$100 K) multiplier stage.  The LO power required at each SIS mixer element is in the range 10 nW to 10 $\mu$W and depends on the frequency and whether several junctions in series are used.
The required phase stability of the LO signal is equivalent to $<12$ fsec of time and the phase noise to $<38$ fsec.

To correct for changes in the electrical length of the fiber, a part of the master laser signal received at each antenna is passed through an acousto-optic cell which shifts the optical frequency by 25 MHz.
The signal is then reflected back down the fiber by a Faraday mirror, which produces a reflected wave for which the plane of polarization is orthogonal to that of the incident wave \cite{Shillue2004}.  At the central facility the phase of this offset frequency is compared with that of the outgoing master laser signal to provide a measure of any variation in the round trip phase out to the antenna and back.  The variation is corrected by means of a fiber stretcher which is adjusted automatically to keep the round-trip phase to a constant value.  A great deal of care has been required to obtain the necessary phase stability of the whole system, including such details as the effect of the variation of the polarization of the two optical signals in the fiber, which requires the use of polarization-maintaining couplers.

Frequency offsets specific to each antenna are required in the first LO signals, to correct for the Doppler shifts resulting from Earth rotation (i.e.\ to remove the fringe-frequency oscillations that would otherwise appear at the correlator output).  For band 10 and the longest baselines, the fringe frequency is $\sim$3 kHz.  Also, a dual phase switching scheme is used.  This consists of phase offsets of 180$^\circ$ which are  helpful in reducing certain spurious responses and possible voltage offsets in the digitization, and phase offsets of 90$^\circ$ for additional sideband separation.  Walsh functions chosen from a set of 128 are used for the phase switching \cite{Emerson2009}.  The two Walsh sets have different time scales: the minimum interval between transitions in the 90$^\circ$ switching is equal to a full Walsh cycle of the 180$^\circ$ switching.  Both the frequency offsets and the phase switching are introduced into the first local oscillator through the phase-locked loop for the YIG-tuned oscillator mentioned above.  The 180$^\circ$ phase offsets can be removed from the signals after they are digitized and in the 12 m array this occurs at the antennas.  The 90$^\circ$ offsets remain until after cross correlation where they can be used for sideband separation for receiver bands in which the first stage is a mixer, see, e.g., \cite{Thompson2001}, or to improve the separation in 2SB front ends.

\section{The ALMA Software}

The development of radio images from visibility data (cross correlations) in radio astronomy requires a number of processing routines that have been developed over several decades, during which improvements in instrumentation have resulted in larger and more accurate observational data bases.  A review of calibration and imaging routines can be found in this issue \cite{Rau2009}.  In particular the narrow antenna beams at the higher frequencies of ALMA will in many cases require multiple pointings, as in mosaicking (forming an image of a wide field from a series of sub-field images).
A comprehensive set of software has been developed for ALMA, including elements which cover the preparation of observing proposals and array schedules, through the operation of the electronic system, to the calibration and pipelined reduction of the data.  A test array, using the prototype antennas,  was operated through Dec.\ 2008 to provide a platform for testing of most elements of the software, including data reduction.  Data reduction is accomplished through use of the Common Astronomy Software Applications (CASA) routines; this software is available for use with other array data.  Ultimately, the ALMA observer will be presented with the raw data, pipeline-reduced images, and the scripts which were used to produce the images, in addition to proposal and scheduling materials.

\section{Summary}

\begin{table}[t]	
\caption{ALMA Sensitivity Goals for the 12-m Array }
\begin{center}
\scriptsize
\begin{tabular}{lccc}  \label{sense}
%\noalign{\medskip}
%\tableline
%\noalign{\smallskip}
Frequency & Continuum$^a$ & Spectral Line$^b$ & Beam, 15 km Configuration \\
  (GHz) & $\Delta$S (mJy$^c$) &  $\Delta$S (mJy) & (arcsec) \\
\hline
\noalign{\smallskip}
110 	&0.047 	&7.0	&  0.038 	\\
140 	&0.055 	&7.1	& 0.030 	\\
230 	&0.100 	&10.2	&0.018 	\\
345 	&0.195 	&16.3	&0.012	\\
409 	&0.296 	&22.6	&0.010	\\
675 	&1.04 	&62.1	&0.006	\\
850 &  1.92       &101         &0.005       \\
\end{tabular}
\end{center}
$^a$Bandwidth = 8 GHz, two polarizations, 64 antennas in one minute.
$^b$Bandwidth = 1 km s$^{-1}$ (equivalent Doppler spread at line frequency), two polarizations.
$^c$One Jansky (Jy) = 10$^{-26}$ W m$^{-2}$ Hz$^{-1}$.
\end{table}

As of this writing, ALMA remains under construction.  An interferometer will operate at the high site for commissioning and astronomical validation near the end of 2009.  During 2011, a phase of Early Science is expected, leading to inauguration of the array during 2012.  The expected sensitivity, based upon specifications, is given for various bands in Table \ref{sense}.  Compared to current capabilities, flux sensitivity and resolution in the millimeter and submillimeter spectral range will increase by almost three orders of magnitude, with an expectation of resultant transformative science.

% or


%\appendices

\section*{Acknowledgment}

None of the progress described in this paper would have been possible without the support of the agencies and institutes listed in footnote 1, together with the sustained efforts of a large number of very talented people who have made possible all of the progress described above.  This paper is presented on behalf of the scientific, engineering, technical and administrative staff of ALMA.  Insightful comments on the manuscript were provided by J. W. M. Baars, A. Baudry, L. D'Addario, D. Emerson, and R. Hills for which we express our appreciation.  Discussions with A. Kerr, P. Napier and D. Thacker were very helpful in developing this presentation of the design of aspects of ALMA.

The National Radio Astronomy Observatory is a facility of the National Science Foundation operated under cooperative agreement by Associated Universities, Inc.

\ifCLASSOPTIONcaptionsoff
  \newpage
\fi



%\bibliographystyle{IEEEtran}
\bibliographystyle{IEEEtran} % Bibliography style file, unsrt.bst
\bibliography{ALMA} % Bibliography database file, moga.bib

\end{document}